
\input amstex
\input vanilla.sty
\nopagenumbers
\baselineskip 14pt
\pagewidth{6in}
\pageheight{8.5in}
\font\ninerm=cmr9
\font\tenrm=cmr10
\font\docerm=cmr12
\TagsOnRight

\def\ov{\overline}
\def\lra{\leftrightarrow}
\def\ovr{\overrightarrow}
\def\noi{\noindent}
{\docerm
\line{\hfil SB/F-93-212}
\vskip 1cm
\centerline{\bf MORE GRAVITATIONAL ANYONS\footnote"*"{\ninerm{To
appear in the proceedings of SILARG VIII, Aguas de Lindoia, July 1993}}}
\vskip 2cm
\centerline{C. Aragone}
\centerline{\it Departamento de F\'{\i}sica, Universidad Sim\'on
Bol\'{\i}var}
\centerline{\it Apartado 89000, Caracas 10800-A,
Venezuela}
\centerline{\it e-mails address: aragone\@ usb.ve}
\centerline{\it and}
\centerline{P. J. Arias}
\centerline{\it Departamento de F\'{\i}sica, Universidad Sim\'on Bol\'{\i}var}
\centerline{\it Apartado 89000, Caracas 10800-A, Venezuela}
\centerline{\it parias\@ usb.ve}
\vskip 2cm
\centerline{\bf ABSTRACT}}
\vskip .3cm

{\tenrm
{\narrower\flushpar
The anyonic behaviour of massive spinning point particles coupled to
linearized massive vector Chern-Simons gravity is studied. This model
constitutes the uniform spin-2 generalization of the vector model formed
by coupling charged point particles to the topological massive Maxwell-CS
action. It turns out that, for this model, the linearized first order triadic
Chern-Simons term is the source of the anyonic behaviour we found.

This is in constrast with the third order topologically massive gravity,
where the anyonic behaviour does not stem in its third-order
Lorentz-Chern-Simons
term, the second order Einstein's  action .\par}}

\newpage

{\docerm
\noi
{\bf 1.	Introduction}
\vskip 4mm

In 2+1 dimensions, particles with arbitrary real spin and non standard
statistics appear naturally. This fact arises, respectively, from the abelian
structure of the two dimensional rotation group, and from the topology of the
many identical particles configuration space$^1$. As a consequence, in 2+1
dimensions, besides the usual ocurrence of bosons and fermions, we can
have anyons$^2$, or particles that have spin and statistics which are neither
bosonic nor fermionic. This ideas can be dynamically implemented when we
consider theories of "bare" bosons (or fermions) with an appropiate
interaction between them. In this theories the particles behave effectively
as ideal anyons i.e. as localized magnetic flux-points$^3$.

This implementation of exotic statistics is commonly done minimally coupling
point particles to electrodynamics with a Chern-Simons term. In these models
the statistics can be understood in terms of the Aharanov-Bohm effect because
the first-order Chern-Simons term induces a magnetic flux-point in  the
position of each particle$^{3,4}$. This picture asymptotically persists  in
the presence of a (second order) Maxwell term.

There is a close analogy between the action of a charged point particle in an
electromagnetic field and the linearized action of a massive point particle
in a weak gravitational field$^5$. So we can define the gravitational analog
of the magnetic flux-point and expect that there should be a dynamical
mechanism to assign them to particles. This was recently proposed$^{6,7}$ in
the context of linearized Topological Massive (TMG) gravity and the anyonic
behaviour found is, for certain circumtances, just like the one obtained for
the corresponding vector theory.

In this note we present a dynamical implementation of exotic statistics in the
context of linearized Massive Vector Chern-Simons (VCSG) gravity. We compute
the anyonic behaviour for a spinning point particle coupled to this linearized
gravity theory. These results are in direct correspondence with those of the
vector analog.

We show that the (linearized) triadic Chern-Simons term works for spin-2
just like the vector Chern-Simons term does for spin-1 models.

\vskip 4mm
\noi
{\bf 2.	Anyonic behaviour in vector models}
\vskip 4mm

The simplest model in which exotic statistics is dynamically implemented
corresponds to consider the quantum mechanics of two identical particles
with the interaction term$^9$
$$
S=<\frac{\theta}{\pi}d\varphi >\tag 1
$$
where $\varphi$ is the relative angle between the two particles, and
$\theta \in [0,\pi ]$ is a numerical parameter, called the statistics
parameter. $\theta$ is 0 or $\pi$ if one is willing to consider, respectively,
two boson or two fermions, and ranges in the open segment, $0<\theta <\pi$,
if one wishes to consider anyons. The induced angular momentum for each
particle will be $\theta /2\pi$.

The interaction term can be rewritten as
$$
S_I=<qd\ovr{r}_{rel}\cdot\ovr{A}(\ovr{r}_{rel})>\ ,\tag 2
$$
with
$$
A_i=-{(q\pi)}^{-1}{\theta}\varepsilon_{ij}\partial_j\ln r_{rel}.
\tag 3 $$
This corresponds to assigning a magnetic flux-point at the origin of the
relative frame. The flux of the "magnetic" field is
$$
\Phi =q^{-1}{2\theta}. \tag 4
$$
In this picture each particle "sees" the other as a magnetic flux-point, and
the Aharanov-Bohm phase will be $q\Phi$.

Let us define the anyonic behaviour parameter by the circulation
$$
\alpha \equiv q\oint dx^iA_i, \tag 5
$$
whose relation with the statistics parameter is $\theta = \alpha /2$.

{}From this naive non relativistic picture we can go futher and consider the
second order action$^6$
$$
S=<-{4}^{-1}F_{mn}F^{mn}-2^{-1}{\mu}\varepsilon^{mnl}A_m\partial_nA_l+
A_mJ^m>,\tag 6
$$
where $F_{mn}=\partial_mA_n-\partial_nA_m$ and $J^m$ is the conserved
particle's source. This action might possess other terms related with the
particles dynamics, here we consider the simplest minded model. Pure
Chern-Simons vector coupling is obtained in the large $\mu$ limit.
Independent variations of $A_l$ in action (6) yield the equations of motion
$$
\partial_mF^{mn}-\mu \varepsilon^{mnl}\partial_nA_l=J^m. \tag 7
$$
Our conventions are $\eta_{mn}=(-++),\varepsilon^{012}=1,
\varepsilon^{oij}\equiv \varepsilon_{ij}$.

We take a circularly symmetric static point current as the source of Eq. 7
$$
J^0=q\delta^{(2)}(\ovr{r})\ ,\ J^i=g\varepsilon_{ij}\partial_j\delta^{(2)}
(\ovr{r}), \tag 8
$$
where we have included, for later comparison, a tranverse spatial dipole
current. In order to obtain $A_r$ we make the anzatz$^6$
$$
A_0=a(r)\ ,\ A_i=\varepsilon_{ij}\partial_jV(r)+\partial_i\lambda .\tag 9
$$
The static solution turns out to be
$$\align
A_0 & = -(q+\mu g)Y(\mu r)\\
A_i & = \varepsilon_{ij}\partial_j(-(\mu^{-1}{q+ g})Y(\mu r)
+\mu^{-1}{q}C(r))\tag 10
\endalign
$$
where $Y(\mu r)$ and $C(r)$ are respectively, the Yukawa and Coulomb Green's
functions
$$
(-\Delta  +\mu^2)Y(\mu r)=\delta^{(2)}(\ovr{r})\ ,\
(-\Delta )C(r)=\delta^{(2)}(\ovr{r}), \tag 11
$$
and
$$
Y(\mu r)=(2\pi)^{-1}K_0(\mu r)\ ,\
C(r)=-(2\pi)^{-1}\ln \mu r. \tag 12
$$

For the pure Chern-Simons theory (wich is the large $\mu$ limit of (7)), the
solutions are
$$
A_0=-\mu^{-1}{g}\delta^{(2)}(\ovr{r})\ ,\
A_i=\mu^{-1}{q}\varepsilon_{ij}\partial_jC(r), \tag 13
$$
for which the anyonic behaviour parameter (5) is
$$
\alpha_{CS}=\mu^{-1}{q^2},\tag 14
$$
where we are thinking in terms of two equal charged particles exchanged. The
dipole current strength $g$ does not affect this result.

For the full theory, with the Maxwell term present, using the fact that the
modified Bessel function behaves asymptotically as
$$
K_0(x)\sim x^{-1/2}e^{-x}, \tag 15
$$
we see that
$$
A_0\sim 0\ ,\ A_i\sim \mu^{-1}{q}\varepsilon_{ij}\partial_jC(r), \tag 16
$$
just as it happens with the pure Chern-Simons solution. Its anyonic behaviour
parameter, for a circle of radius $R$, is
$$
\alpha_{TM}(R)=\mu^{-1}{q^2}+{q}(q+\mu g) RK_1(\mu R), \tag17
$$
which differs from (14) unless $q+\mu g=0$. Asymptotically the two anyonic
behaviours for both theories coincide, as expected, due to the fact that the
first-order Chern-Simons action is the term which provides the dominant
contribution in this limit.

\vskip 4mm
\noi
{\bf 3.	Anyonic behaviour for linearized MVCS gravity}
\vskip 4mm

The possibility of implementing exotic statistics in linearized gravity can
be studied by considering the action of massive monopolar point particle
moving in an external gravitational field
$$
S_p=-m\int d\tau (-g_{mn}\dot{x}^m\dot{x}^n)^{1/2}\tag 18
$$
and then taking the limit for small velocities , going to the weak
field  approximation $g_{mn}=\eta_{mn}+\kappa h_{\ov{mn}}$,
$h_{\ov{mn}}=h_{\ov{nm}}$. In this approximation the hamiltonian density
takes the form$^5$
$$
H={(2m)}^{-1}{(\ovr{p}-q\ovr{A})}^{2}+qV,\tag 19
$$
with
$$
qA_i=\kappa mh_{\ov{0i}}\ ,\
qV=-{2}^{-1}\kappa mh_{\ov{00}}.\tag 20
$$

Notice that the expression for $H$ in Eq. 19 corresponds to the
hamiltonian density of a charged particle in an electromagnetic field whose
vector potential is $A_\mu =(V,A_i)$. In this weak field approximation the
linear coupling with matter can be seen easily to be
$$
S_C=2^{-1}{\kappa m}<h_{\ov{mn}}T^{mn}>,\tag 21
$$
where $T^{mn}$ is the particle's flat energy momentum tensor. If we introduce
the dreibeins $e_m{}^a$ with $g_{mn}=e_m{}^ae_n{}^b\eta_{ab}$, and make
the linearization $e_m{}^a=\delta_m{}^a+\kappa h_m{}^a$, then
$$
h_{\ov{mn}}=h_{mn}+h_{nm}, \tag 22
$$
($h_{mn}$ is not symmetric) and the linear coupling with matter that arises
will be $\kappa mh_{mn}$ $T^{mn}$ with $T^{mn}=T^{nm}$ as in (21).

We now consider the minimally coupled action
$$\align
S_2=
&{2}^{-1}<h_{pa}({2}^{-1}\varepsilon^{pma}\varepsilon^{srb}-\varepsilon^{pmb}
\varepsilon^{sra})\partial_m\partial_rh_{sb}>+\\
&+2^{-1}{\mu}<h_{pa}\varepsilon^{prs}\partial_rh_s{}^a>+\kappa <h_{mn}T^{mn}>\\
\equiv & S_E^L+S^L_{TCS}+S_C, \tag 23
\endalign
$$
where $S_E^L+S_{TCS}^L$ is the quadratic approximation of the curved
second-order
action of massive vector Chern-Simons gravity$^8$
$$
S_{MVCS}={(2\kappa^2)}^{-1}<e_{pa}\varepsilon^{pmn}R_{mn}^{*a}(\omega (e))+ \mu
e_{pa}\varepsilon^{pmn}\partial_me_n{}^a>, \tag 24
$$

The first term in (24) is Einstein's action with
$R^{*a}_{mn}=\partial_m\omega_n^a-\partial_n\omega_m^a-\varepsilon^a_{bc}
\omega_m^b\omega_n^c$ where $\omega_m^a(e)$ is the value of the torsionless
connection given in terms of the dreibeins i.e. such that
$\varepsilon^{pmn}(\partial_me_n^a-\varepsilon^a{bc}\omega_m^be_n^c)=0$. The
second term in Eq. 24 is the Triadic Chern-Simons (TCS) term and its
quadratic part, as we will show, constitutes the gravitational analog of the
Chern-Simons vector term which allow the dynamical implementation of
statistics.

$S_E^L+S^L_{TCS}$ was introduced as an intermediate action which interpolates
between first-order self-dual gravity and the master action for linearized
gravity theories$^{10}$. It propagates one excitation of helicity
$\pm 2\mu /|\mu |$, depending on the sign of $\mu^8$. This action is
invariant under the gauge transformations $\delta h_{mn}=\partial_m\xi_n$ and
is equivalent, as a free theory, both to self-dual gravity and to linearized
TMG$^{10,11}$ This equivalence will not persists, as it will be shown, in this
interaction picture, if the theories are coupled to point particles under the
same footings. So let us pass to the dynamical analysis.

The equation of motion which follows from (23) is
$$
({2}^{-1}\varepsilon^{pml}\varepsilon^{srn}-\varepsilon^{pmn}
\varepsilon^{srl})\partial_m\partial_rh_{sn}+\mu\varepsilon^{psr}\partial_s
h_r{}^l=-kT^{pl}. \tag 25
$$
We take the static source
$$
T^{00} =m\delta^{(2)}(\ovr{r})\ ,\
T^{0i}=2^{-1}{\sigma}\varepsilon{ij}\partial_j\delta^{(2)}(\ovr{r})\ ,\
T^{ij}=0,\tag 26
$$
wich corresponds to a massive ($\int T^\infty d^2x=m$), spinning
($\int \varepsilon_{ij}x^iT^{0j}d^2x=\sigma$) point particle. Looking for
static solutions, we choose the natural $T+L$ decomposition (after projection
of
the unsymmetric linearized dreibein into its symmetric plus antisymmetric parts
$h_{mn}={2}^{-1}h_{\ov{mn}}+\epsilon_{mnl}v^l$)
$$\align
h_{00} & = n(r)\\
h_{0i} & = \varepsilon_{ij}\partial_j(n^T-V^L)+\partial_i(n^L+V^T)\\
h_{i0} & = \varepsilon_{ij}\partial_j(n^T-V^L)+\partial_i(n^L+V^T)\\
h_{ij} & = (\delta_{ij}\Delta -\partial_i\partial_j)h^T+
(\varepsilon_{ik}\partial_k\partial_j+\varepsilon_{jk}\partial_k\partial_i)
h^{TL}+\partial_i\partial_jh^L+\varepsilon_{ij}V. \tag 27
\endalign
$$

We take the gauge $V=V^T=0$, $h^L=h^T$ and insert the structures (26) (27)
into Eq. 25 in order to determine $h^T$, $h^{TL}$, $n^T$, $n^L$, $n$.

We obtain for the linearized metric ($h_{\ov{mn}}=h_{mn}+h_{nm}$)
$$\align
h_{\ov{00}} & = \kappa (m+\mu\sigma )Y(\mu r),\\
h_{\ov{0i}} & = -\varepsilon_{ij}\partial_j ({\kappa }({\mu}^{-1}m+\sigma )
Y(\mu r)+\mu^{-1}{\kappa m}C(r)), \\
h_{\ov{ij}} & = \delta_{ij}\kappa (m+\mu\sigma )Y(\mu r).\tag 28
\endalign
$$

The large $\mu$ limit of (25) represents pure TCS theory. Its solutions are,
for the linearized metric
$$
h_{\ov{00}}=\mu^{-1}{\kappa \sigma}\delta^{(2)}(\ovr{r})\ ,\
h_{\ov{0i}}=\mu^{-1}{\kappa m}\varepsilon_{ij}\partial_jC(r)\ ,\
h_{\ov{ij}}=\mu^{-1}{\kappa \sigma }\delta_{ij}\delta^{(2)}(\ovr{r}). \tag 29
$$

It is the gravitational generalization of (13) provided we make the
identification
$$
\kappa m\lra q\ ,\ \kappa \sigma \lra g\ ,\ h_{00}\lra -A_0\ ,\
h_{0i}\lra A_i. \tag 30
$$

The anyonic behaviour parameter is
$$
\alpha_{TCS}=\kappa m\oint dx^ih_{\ov{0i}}=
\mu^{-1}{\kappa^2 m^2}, \tag 31
$$
independently of any contour, as expected.

For the, coupled, linearized MVCS gravity theory, we note that the static
solutions, (28) correspond to the generalization of (10) with the above
mentioned identifications. The asymptotic behaviour of the solutions is
$$
h_{\ov{00}}\sim 0\ , \
h_{\ov{0i}}\sim -(2\pi\mu)^{-1}\kappa m\varepsilon_{ij}\partial_j\ln r\ , \
h_{\ov{ij}}\sim 0,
$$
confirming this assertions. Its anyonic behaviour parameter for a circle
of radius $R$, as expected from (17), is given by
$$
\alpha_{MVCS}(R)=\mu^{-1}(\kappa^2 m^2)+({\kappa m})\kappa
(m+\mu\sigma) R K_1(\mu R).\tag 32
$$
For the special case that $m+\mu\sigma =0$ it is identical to Eq. 31. It also
asymptotically coincides with it.

We see, then, that massive point particles independently of its intrinsic
angular momentum look like ``flux-points" when the interaction term between
them is a linearized TCS term. This picture persists at finite distances and
also,  if the linearized Einstein term is present, in the special case that
$m+\mu\sigma =0$. For any other value of $\sigma$ the results remains true
asymptotically. So massive point particles behave as anyons with statistic
parameter $2^{-1}\mu^{-1}\kappa^2m^2$.

\vskip 4mm
\noi
{\bf 4.	Conclusions}
\vskip 4mm

We presented a model that dynamically implements anyonic statistics in the
context of linearized gravity models. Coupling massive spinning point
particles with a linearized TCS term constitute the exact analog of the pure
Chern-Simons vector theory for charged particles. In the former model massive
particles behave as anyons with statistic parameter $2^{-1}\mu^{-1}\kappa^2m^2$
and induced  angular momentum $(4\pi\mu)^{-1}\kappa^2m^2$.

The analogy goes over in the presence of Einstein's term. Linearized massive
vector-CS gravity is the uniform spin-2 generalization of Maxwell-CS theory.
This is not the case for linearized topological massive gravity$^{6,7}$,
because its linearized version only works as an analog of MCS if there is
no intrinsic angular momentum $\sigma$, and no dipole current strenght $g$.
In the case that $m+\mu\sigma =0$ there is no induced angular momentum
for TMG. In contrast, for the MVCS model, it behaves just as the pure
TCS model, exactly as it happens when you consider the vector analog with
$q+\mu g=0$ and compare with the pure Chern-Simons theory.

We expect that the calculus of the induced spin from the generators of
rotations will have close similarities with the vector model calculations$^6$.
These calculations are now in progress and will be reported elsewhere.

\vskip 4mm
\noi
{\bf \item{5.}References}
\vskip 4mm

\item{1.}J. M. Leinaas and J. Myrheim, Nuovo Cimento {\bf B37} (1977) 1; B.
Binegar, J. Math. Phys. {\bf 23} (1982) 1511.
\item{2.}F. Wilczek, Phys. Rev.Lett. {\bf 48} (1982); ibid. {\bf 49}  (1982)
957.
\item{3.}P. Arovas, R. Schrieffer, F. Wilczek and A. Zee, Nucl. Phys.
{\bf B251} (1985) 117; R. Mackenzie and F. Wilczek, Int. J. Mod. Phys. {\bf A3}
(1988) 2827,  and references there in.
\item{4.}S. Rao, {\it An anyon premier}, Preprint TIFR/TH/92-18; A. Khare, {\it
Quantum Mechanics and Statistics Mechanics of Anyons}, in  the
centenary issue of Holkar Science College, Indore, India.
\item{5.}B. De Witt, Phys. Rev. Lett. {\bf16} (1966) 1092.
\item{6.}S. Deser,Phys. Rev. Lett. {\bf 64} (1990) 611; S. Deser and J. G.
McCarthy, Nucl. Phys. {\bf B344} (1990) 747; S. Deser, Class. Quantum Grav.
{\bf 9} (1992) 61.
\item{7.}M. E. Ortiz, Nucl. Phys. {\bf B} (1991).
\item{8.}C. Aragone, P. J. Arias and A. Khoudeir, {\it Massive Vector
Chern-Simons gravity}, Pre\-print SB/F-192/92.
\item{9.}Mackenzie and F. Wilczek, in ref. 3.
\item{10.}C. Aragone and A. Khoudeir,
Phys. Lett. {\bf B173} (1986) 141; C. Aragone and A. Khoudeir, {\it
Quantum Mechanics of Fundamental  Systems}, eds. C. Teitelboim and J. Zanelli.
p.
17.
\item{11.}S. Deser and J. G.McCarthy, Phys. Lett. {\bf B245} (1990)
441; (Addendum) {\bf B248} (1990) 473.
 \bye